\documentclass[12pt]{article}

\usepackage{slashed}

\advance \textheight by \topskip
\makeatletter
\@addtoreset{equation}{section}
 
\makeatother

\newcommand{\be}{\begin{equation}}
\newcommand{\ee}{\end{equation}}
\newcommand{\bea}{\begin{eqnarray}}
\newcommand{\eea}{\end{eqnarray}}
\newcommand{\dd}{\partial}

\newcommand{\dirac}{\partial\hskip-0.22truecm /}

\newcommand{\gauge}{A\hskip-0.22truecm /}

\def\>{\rangle}
\def\<{\langle}

\begin{document}

\title{
{\bf Gauge invariance and non-constant gauge couplings}}


\author{
{\sf   N. Mohammedi} \thanks{e-mail:
nouri@lmpt.univ-tours.fr}$\,\,$${}$
\\
{\small ${}${\it Laboratoire de Math\'ematiques et Physique Th\'eorique (CNRS - UMR 6083),}} \\
{\small {\it F\'ed\'eration Denis Poisson (FR CNRS 2964)}},\\
{\small {\it Universit\'e Fran\c{c}ois Rabelais de Tours,}}\\
{\small {\it Facult\'e des Sciences et Techniques,}}\\
{\small {\it Parc de Grandmont, F-37200 Tours, France.}}}
\date{}
\maketitle
\vskip-1.5cm

\vspace{2truecm}

\begin{abstract}

\noindent
It is shown that space-time dependent gauge couplings do not completely break gauge invariance.  
We demonstrate this in various gauge theories.

\end{abstract}

\newpage

%

\setcounter{equation}{0}

\section{Introduction}

Gauge invariance is certainly one of the most important guiding 
principles in modern particle physics. 
Among its key features
is that it prevents the gauge fields from acquiring a direct mass term. 
However, this obstacle
can be circumvented through the Higgs mechanism \cite{higgs,brout,kibble} 
where the mass of the gauge field
is a space-time dependent quantity governed by a scalar field.
When this scalar field falls into its ground state, 
the mass term becomes constant. 
\par
Another characteristic (or common belief) of gauge invariance is that it forces 
the gauge couplings to be constant. In this article, we would like to ask
the question of whether gauge invariance is completely lost if 
the gauge couplings are space-time dependent quantities. The conclusion of this investigation 
is that some gauge invariance is still present. 
\par
The issue of non-constant gauge couplings arises, at least,  in two contexts.
The first is encountered in the procedure of renormalisation where at the quatum level
the couplings are function of the energy scale (running couplings) or equivalently
functions of distance. {} For example, in quantum electrodynamics the 
gauge coupling increases with energy (that is, with decreasing distance).
Yet this quantum phenomenon is completely absent in the classical theory.
It would, therefore, be  desirable to see if this property 
(space-time dependence) of the gauge couplings 
could be implemented at the classical level.  
\par
The second domain where non-constant gauge couplings could be of relevance
is in cosmology and astrophysics. Indeed, one might reasonably challenge the
assumption  that 
the electric charge (the gauge coupling of quantum electrodynamics),
or other gauge couplings in non-Abelian gauge theories,  
should be constant at all times and in all regions of the Universe.

\par
In this context, 
theories modelling the space-time dependence of some gauge couplings
have already appeared in the literature \cite{bekenstein,barrow1,barrow2,olive1,olive2}. 
A general review of the subject could also be found in \cite{uzan}.
They are based on the
introduction of  new scalar fields and their essence could be summarised by the 
Lagrangian \cite{bekenstein}
\bea
{\cal L} &=& -{1\over 4}e^{-2\varphi}F_{\mu\nu}F^{\mu\nu} +\bar \psi\left[i\dirac
-e\gauge -m\right]\psi 
+{1\over 2}\dd_\mu\varphi \dd^\mu\varphi -V\left(\varphi\right)
\nonumber\\
F_{\mu\nu} &=&\dd_\mu A_\nu -\dd_\nu A_\mu \,\,\,\,\,.
\eea
Of course when $\varphi=0$ (and assuming that $V(0)=0$), 
this Lagrangian reduces to the usual field theory of quantum electrodynamics{\footnote{The
fermionic contribution $\bar \psi\left[i\dirac
-e\gauge -m\right]\psi$ is not present in ref.\cite{bekenstein}.}}.  
We have, for simplicity, omitted the gravity sector.  

\par
 
This theory is invariant under the gauge transformation 
\be
\psi\,\longrightarrow\,e^{-ie\alpha(x)}\,\psi\,\,\,\,\,\,\,\,\,,\,\,\,\,\,\,\,\,     
A_\mu\,\longrightarrow\, A_\mu + \dd_\mu \alpha(x)\,\,\,\,\,\,\,
\ee
and 
the equations of motion corresponding to the gauge field $A_\mu$,
the fermion field $\psi$ and the scalar field $\varphi$
are, respectively,  given by
\bea
 &&\dd_\mu F^{\mu\nu} = e\,e^{2\varphi}\bar\psi\gamma^\nu\psi + 2\dd_\mu\varphi F^{\mu\nu} 
 \nonumber\\
&&\left[i\dirac
-e\gauge -m\right]\psi =0
\nonumber\\
&&
\dd^\mu\dd_\mu\varphi= -{dV\over d\varphi}+{1\over 2}e^{-2\varphi}F_{\mu\nu}F^{\mu\nu}
\,\,\,\,\,.
\label{asymm}
\eea
We immediately notice that if one sets the fermion field $\psi$ to zero then what remains 
of the first equation in (\ref{asymm}) is 
\be
\dd_\mu F^{\mu\nu} = 2\dd_\mu\varphi F^{\mu\nu}\,\,\,\,\,.
\ee
Clearly, these are not Maxwell's equations in the vaccum ($\dd_\mu F^{\mu\nu}=0$).
Hence, the scalar field is present even if
the electromagnetic interaction between the gauge field $A_\mu$ and the charged fermion
$\psi$ is absent. 
Furthermore, there is an `asymmetry` between 
the first two equations of (\ref{asymm}). In the first equation, the coupling to the 
electromagnetic current $\bar\psi\gamma^\nu\psi$ is $e\,e^{2\varphi}$ while 
the electromagnetic coupling in the Dirac equation (the second equation) 
is $e$.

\par
In this paper we adopt the strategy of taking the existing
gauge theories and simply replace the constant
gauge couplings by non-constant ones and demand that gauge invariance holds.
{}For instance, in the case of quantum electrodynamics,  
the electromagnetic interaction $e\,\bar\psi\gauge\psi$ (with the constant 
gauge coupling $e$) will be replaced by $\widetilde e\left(x\right)\,\bar\psi\gauge\psi$.
In this way,  one has the same interaction terms (or vertices) between the
gauge fields and the other fields 
as in the case of constant gauge couplings.
We start by examining the case 
a relativistic charged particle interacting with an electromagnetic
field. We then extend the analyses to quantum mechanics, quantum electrodynamics,
non-Abelian gauge theories, the Abelian Higgs model  and the electro-weak theory.


\section{Coupling of a charged point particle to the electromagnetic field}

The usual Lagrangian for a relativistic charged particle 
of mass $m$ interacting with an electromagnetic
field $(\vec{E}\,,\,\vec{B})$ is (see for instance \cite{landau})
\be
{\cal {L}}=-mc^2\sqrt{1-{v^2\over c^2}} + {e\over c} \vec{A}.\vec{v}
-e\varphi\,\,\,\,.
\ee 
The strength of this coupling is the constant $e$ while  
$c$ is the speed of light. The velocity vector $\vec{v}={d\vec{r}\over dt}$
with $\vec{r}(t)$ being the vector position of the particle.
\par
The vector potential $\vec{A}$ 
and the scalar potential $\varphi$ are the quantities in terms of which 
the electric field $\vec{E}$ and the magnetic field $\vec{B}$ are defined.
These are given by
\bea
\vec{E} &=& -{1\over c}{\dd \vec{A}\over \dd t}-\vec{\nabla}\varphi
\nonumber\\
\vec{B} &=& \vec{\nabla}\wedge\vec{A}
\eea
The electric and magnetic fields are invariant under
\bea
\vec{A}\longrightarrow \vec{A}-\vec{\nabla}\alpha
\,\,\,\,\,\,\,\,\,,\,\,\,\,\,\,\,\,
\varphi\longrightarrow \varphi +{1\over c}{\dd \alpha\over \dd t}\,\,\,\,,
\label{gauge1}
\eea
where $\alpha=\alpha(\vec{r}\,,\,t)$ is an arbitrary function.
Under this gauge transformation the Lagrangian transforms as
\be
{\cal{L}}\longrightarrow {\cal{L}} -{e\over c}{d\alpha\over dt}\,\,\,\,\,.
\ee
The additional term $-{e\over c}{d\alpha\over dt}$ is a total derivative in time.
Hence, the action $S=\int_{t_{1}}^{t_{2}}{\cal {L}}\,dt$ 
changes by a constant and the equations of motion are, as a consequence, invariant. 
\par
Suppose now that the strength of the coupling between the charged
particle and the electromagnetic field is not constant. Namely, we consider the 
Lagrangian
\be
{\cal {L}}=-mc^2\sqrt{1-{v^2\over c^2}} + {{\widetilde{e}}\over c} \vec{A}.\vec{v}
-{\widetilde{e}}\varphi\,\,\,\,,
\label{CED}
\ee 
where the coupling ${\widetilde{e}}$ is a function of
space and time. That is, ${\widetilde{e}}={\widetilde{e}}(\vec{r}\,,\,t)$.
The speed of light $c$ is assumed, through out this paper,  to be constant.

\par
Since the definitions of the $\vec E$ and $\vec B$ have not changed,
the gauge symmetry is still as in (\ref{gauge1}).
Under the gauge transformation (\ref{gauge1}) the Lagrangian transforms again as
\be
{\cal{L}}\longrightarrow {\cal{L}} -{{\widetilde{e}}\over c}{d\alpha\over dt}\,\,\,.
\ee
In order for gauge invariance to hold at the level of the  
action $S=\int_{t_{1}}^{t_{2}} {\cal {L}}\,dt$, 
we demand that this variation is a total differential in time. 
That is
\be
{{\widetilde{e}}\over c}
{d\alpha\over dt}
={d\beta\over dt} \,\,\,\,\,\,
\ee
for some function $\beta$. 
This requirement is fulfilled if
the gauge parameter $\alpha$ is an arbitrary function of ${\widetilde{e}}(\vec{r}\,,\,t)$, 
namely $\alpha=\alpha({\widetilde{e}})$,
and in this case we have
\be
\beta({\widetilde{e}})=\int\,{{\widetilde{e}}\over c}{d\alpha\over d{\widetilde{e}}}\,d{\widetilde{e}}
\ee
We conclude that even if the coupling ${\widetilde{e}}$ is not constant, gauge invariance is not completely
lost.
\par
The Lagrangian (\ref{CED}) leads to the following equations of motion   
\bea
{{\rm d}\vec{p}\over {\rm d}t} &=& {{\rm d}\over {\rm d}t}\left({m\vec{v}\over\sqrt{1-{v^2\over c^2}}}\right)
= {\widetilde{e}}\vec{E} +{{\widetilde{e}}\over c}\vec{v}\wedge \vec{B}
+\vec{F}_{{\widetilde e}}
\nonumber \\
\vec{F}_{{\widetilde e}} &=& {1\over c}\left(\left(\vec{A}.\vec{v}\right)-c\varphi\right)
\vec\nabla {\widetilde{e}}
-{1\over c}\left(\left(\vec{v}.\vec\nabla {\widetilde{e}}\right) 
+{\dd {\widetilde{e}}\over \dd t}\right)\vec{A}
 \,\,\,\,\,.
\eea
The forces acting on the particle are  the Lorentz
force (but with a space-time dependent electric charge $\widetilde{e}$)
plus another force $\vec{F}_{{\widetilde e}}$
due to the fact that the gauge coupling
$\widetilde{e}$ is not constant.  
\par
One might be tempted to simply redefine the vector potential  $\vec{A}$ as 
$\vec{A}\longrightarrow {e\over{\widetilde{e}}} \vec{A}$ and the scalar
potentiel $\varphi$ as $\varphi\longrightarrow {e\over {\widetilde{e}}}\varphi$,
where $e$ is a constant, 
in the Lagrangian (\ref{CED}) 
in order to absorb the space-time dependence
of the coupling ${\widetilde{e}}$. This is indeed possible if the electromagnetic field
is not dynamical. In the full theory, however,    
the gauge invariant Lagrangian for the electromagnetic field is as usual given by
\be
{\cal L}_{{\rm gauge}}={1\over 8\pi}\int\left(E^2-B^2\right)\,dV\,\,\,\,,
\ee
where $dV=dx\,dy\,dz$ is the volume element. Hence a redefinition of the gauge fields
would induce a change in the expressions of $\vec{E}$ and $\vec{B}$ which results in  
a non-standard kinetic term for the electromagnetic field. 
\par
Maxwell's equation stem from the variation 
with respect to $\vec{A}$ and $\varphi$ of the full action
\be
S=\int\left[-mc^2\sqrt{1-{v^2\over c^2}} + {{\widetilde{e}}\over c} \vec{A}.\vec{v}
-{\widetilde{e}}\varphi \right]dt 
+ {1\over 8\pi}\int\left(E^2-B^2\right)dVdt
\,\,\,\,.
\ee
In order to introduce the concept of the charge density, let us write 
\be
{\widetilde{e}}(\vec{r}\,,\,t)=e\,\lambda(\vec{r}\,,\,t)
\ee
with $e$ a constant (to be identified with the charge of the point particle).
The charge density and the current density are then defined by 
\be
e= \int \rho \,dV\,\,\,\,\,\,\, {\mathrm{with}}\,\,\,\,\,\,\,\,  
\rho=e\,\delta\left(\vec{r}-\vec{r}_0\right)
\,\,\,\,\,\,\, {\mathrm{and }}\,\,\,\,\,\,\,\,  
\vec{{\jmath}}=\rho\,\vec{v} \,\,\,\,,
\label{j}
\ee
where $\vec{r}_0$ is the vector position of the charge $e$
whose vector velocity is $\vec{v}$. 
\par
In this way the action 
becomes
\be
S=\int\left(-mc^2\sqrt{1-{v^2\over c^2}}\right)dt 
+\int\left[{\lambda\over c}\rho\left(\vec{A}.\vec{v}
-c\varphi \right) 
+ {1\over 8\pi}\int\left(E^2-B^2\right)\right]dVdt
\,\,\,\,.
\ee 
The field equations are found by demanding that $\delta S=0$ 
under the variations $\vec{A}\longrightarrow \vec{A} +\delta\vec{A}$
and $\varphi\longrightarrow \varphi +\delta\varphi$ (assuming, of course,
that the motion of the charge is known). This procedure yields
\bea
\vec{\nabla}.\vec{E} &=& 4\pi\lambda\rho
\nonumber \\
\vec{\nabla}\wedge\vec{B} &=&{1\over c}{\dd \vec{E}\over \dd t}
+{4\pi\over c}\lambda \, \vec{{\jmath}}\,\,\,\,.
\eea
A continuity equation (conservation equation) is established by 
taking the divergence of the second equation (
$\vec\nabla.(\vec{\nabla}\wedge\vec{B})=0$). This is given by
\bea
{\dd\over \dd t}\left(\lambda\rho\right)+\vec\nabla.\left(\lambda\,\vec{{\jmath}}\right)=0
\,\,\,\,.
\eea
Replacing $\rho$ and $\vec j$ by their expressions in (\ref{j})
and $\lambda$ by $\widetilde e/e$ yields
\bea
{\dd {\widetilde{e}}\over \dd t}
+\left(\vec{v}.\vec\nabla {\widetilde{e}}\right) 
=0
 \,\,\,\,\,.
\eea
Hence the extra force $\vec{F}_{{\widetilde e}}$ takes the form
\bea
\vec{F}_{{\widetilde e}} &=& {1\over c}\left(\left(\vec{A}.\vec{v}\right)-c\varphi\right)
\vec\nabla {\widetilde{e}}
\,\,\,\,\,.
\eea


\section{Quantum mechanics}

The Schr\"{o}dinger equation for a non-relativistic charged particle moving
through an electromagnetic field is (see for instance \cite{schiff})
\bea
i\hbar {\dd \psi\over \dd t}=\left[{1 \over 2m}
\left(-i\hbar \vec{\nabla}
-{e \over c}\vec{A}\right)^2 +e\varphi\right]\psi \,\,\,\,.
\label{shro1}
\eea
This equation is invariant under the gauge transformation
(\ref{gauge1}) if the wave function transforms as
\be
\psi \longrightarrow e^{-{ie\alpha\over\hbar c}}\,\psi\,\,\,\,\,.
\ee
As a consequence, the probability density 
and the probabilty current are invariant.
\par 
Let us now assume that the strength of the gauge interaction, $e$, 
is no longer a constant and consider, instead,  the equation
\bea
i\hbar {\dd \psi\over \dd t}=\left[{1 \over 2m}
\left(-i\hbar \vec{\nabla}
-{\widetilde {e} \over c}\vec{A}\right)^2 +\widetilde{e}\varphi\right]\psi \,\,\,\,,
\label{shro2}
\eea
where $\widetilde{e}={\widetilde{e}}(\vec{r}\,,\,t)$. We demand that this equation
is still invariant under 
\bea
\vec{A}\longrightarrow \vec{A}-\vec{\nabla}\alpha
\,\,\,\,\,\,\,\,\,,\,\,\,\,\,\,\,\,
\varphi\longrightarrow \varphi +{1\over c}{\dd \alpha\over \dd t}
\,\,\,\,\,\,\,\,\,,\,\,\,\,\,\,\,\,
\psi \longrightarrow e^{-i\beta}\psi
\eea
for some $\alpha$ and $\beta$ to be determined.  In order for the
above equation to remain invariant, one must have
\be
\vec{\nabla}\beta={\widetilde{e}\over \hbar c}\vec{\nabla}\alpha\,\,\,\,\,,\,\,\,\,\,\,
{\dd \beta\over \dd t}= {\widetilde{e}\over \hbar c}{\dd \alpha\over \dd t}\,\,\,\,\,.
\ee
These conditions have a solution if $\alpha$ is an arbitrary function of $\widetilde{e}$,
that is $\alpha=\alpha(\widetilde{e})$, and $\beta$ is given by
\be
\beta({\widetilde{e}})=\int\,{{\widetilde{e}}\over \hbar c}{d\alpha\over d{\widetilde{e}}}\,d{\widetilde{e}}
\,\,\,\,.
\ee
We deduce here also that some gauge invariance is still present in 
the Schr\"{o}dinger equation (\ref{shro2})
with non-constant electromagnetic coupling.

\section{Quantum electrodynamics}

Quantum electrodynamic is a theory describing the interaction of charged fields
with radiation. In the case of fermionic fields, the theory is
given by the classical Lagrangian (see for instance \cite{ryder,quigg,zuber})
\bea
{\cal L} &=& -{1\over 4} F_{\mu\nu}F^{\mu\nu} +\bar \psi\left[i\dirac 
-e\gauge -m\right]\psi \nonumber\\
F_{\mu\nu} &=&\dd_\mu A_\nu -\dd_\nu A_\mu
\label{QED}
\eea
When the coupling
$e$ between the fermion field $\psi$ and the radiation field $A_\mu$ is constant, 
the theory is invariant under the  gauge transformation
\be
\psi\,\longrightarrow\,e^{-ie\alpha(x)}\,\psi\,\,\,\,\,\,\,\,\,,\,\,\,\,\,\,\,\,     
A_\mu\,\longrightarrow\, A_\mu + \dd_\mu \alpha(x)\,\,\,\,\,\,\,,
\label{gaugetransf1}
\ee
where $\alpha(x)$ is a completely arbitrary function of the space-time coordinates.
\par
Let us now examine what becomes of this gauge invariance if the gauge coupling
is a space-time dependent function . The Lagrangian is of the same form as before
\bea
{\cal L} &=& -{1\over 4}F_{\mu\nu}F^{\mu\nu} +\bar \psi\left[i\dirac
-{\widetilde{e}}\gauge -m\right]\psi \nonumber\\
F_{\mu\nu} &=&\dd_\mu A_\nu -\dd_\nu A_\mu
\label{qed}
\eea
but now ${\widetilde{e}}={\widetilde{e}}(x)$. 
In analogy with (\ref{gaugetransf1}), we demand that 
this Lagrangian is invariant under
\be
\psi\,\longrightarrow\,e^{-i\beta(x)}\,\psi\,\,\,\,\,\,\,\,\,,\,\,\,\,\,\,\,\,     
A_\mu\,\longrightarrow\, A_\mu + \dd_\mu \alpha(x)\,\,\,\,
\label{gaugetransf2}
\ee
for some functions $\alpha(x)$ and $\beta(x)$. It follows 
that the Lagrangian (\ref{qed}) remains invariant under the
transformations (\ref{gaugetransf2})
provided that $\alpha(x)$ and $\beta(x)$ satisfy
\be
\dd_\mu\beta(x) ={\widetilde{e}}(x)\, \dd_\mu\alpha(x)
\,\,\,\,\,\,\,\,{\rm{or}}\,\,\,\,\,\,\,\,
d\beta={\widetilde{e}}\,d\alpha
\,\,\,\,\,.
\ee
This equation is consistent only if
\be
\dd_\mu {\widetilde{e}} \,\dd_\nu\alpha
=\dd_\nu {\widetilde{e}} \,\dd_\mu\alpha
\,\,\,\,\,\,\,\,{\rm{or}}\,\,\,\,\,\,\,\,
d{\widetilde{e}}\wedge\,d\alpha=0
\,\,\,\,\,.
\ee
This last condition has a solution if 
$\alpha(x)=\alpha({\widetilde{e}}(x))$.
That is,  $\alpha$ is an arbitrary function of the coupling ${\widetilde{e}}(x)$. In this case
$\beta$ is also a function of $\widetilde{e}$ and is given by{\footnote{
We could also have written $\beta=\int{\widetilde{e}}\,d\alpha$. This means 
that ${\widetilde{e}}$ is given in terms of $\alpha$. 
However, one is first given a Lagrangian (that is ${\widetilde{e}}$)
and then one looks for the symmetries of this Lagrangian (that is $\alpha$).
This is why we prefer to say that the gauge function $\alpha$ is
expressed in terms of the coupling  ${\widetilde{e}}$. }}
\be
\beta({\widetilde{e}}(x)) = \int \left({\widetilde{e}} {d\alpha\over d{\widetilde{e}}}\right)\, 
d{\widetilde{e}} 
\,\,\,\,\,\,\,\,.
\ee
This last equation can be written, after an integration by parts, as
\be
\beta={\widetilde{e}}\,\alpha - \int \alpha\, d{\widetilde {e}} 
\,\,\,\,\,\,\,\,.
\ee
We can clearly  see that if ${\widetilde {e}}$ is independent of the space-time
points (that is, $d{\widetilde {e}}=0$) then $\beta={\widetilde{e}}\,\alpha$
and the transformations (\ref{gaugetransf2}) are the usual 
gauge transformations of ordinary quantum electrodynamics with a constant gauge coupling.
To summarise, the Lagrangian (\ref{qed}), with ${\widetilde{e}}={\widetilde{e}}(x)$, is invariant under 
the local transformations
\be
\psi\,\longrightarrow\,
e^{-i \int \left({\widetilde{e}} {d\alpha\over d{\widetilde{e}}}\right)\,d{\widetilde{e}}}\,\psi
\,\,\,\,\,\,\,\,\,,\,\,\,\,\,\,\,\,     
A_\mu\,\longrightarrow\, A_\mu + \dd_\mu \alpha
\,\,\,\,,
\ee
where $\alpha({\widetilde{e}}(x))$ is arbitrary.
\par
The equations of motion corresponding to the Lagrangian (\ref{qed}) are
\bea
 &&\dd_\mu F^{\mu\nu} = {\widetilde{e}}\,\bar\psi\gamma^\nu\psi 
 \nonumber\\
&&\left[i\gamma^\mu\dd_\mu 
-{\widetilde{e}}\,\gamma^\mu A_{\mu} -m\right]\psi =0
\eea
The first equation implies that the current
\bea
J^\nu &=& {\widetilde{e}}\bar\psi\gamma^\nu\psi 
\eea
is conserved ($\dd_\nu J^\nu =0$). 
The corresponding conserved charge is
\be
Q=\int  J^0d^3x= \int {\widetilde{e}}\bar\psi\gamma^0\psi\, d^3x
=\int {\widetilde{e}}\psi^\dagger\psi \,d^3x \,\,\,\,.
\ee
As ${\widetilde{e}}(x)$ cannot be taken out of the integral,
the conserved quantity $Q$ is the integral  of the 
fermionic `probability density`   $\widetilde{e}\psi^\dagger\psi
=(\sqrt{\widetilde{e}}\psi)^\dagger(\sqrt{\widetilde{e}}\psi)$.
It is as if the `wave function` is $\sqrt{\widetilde{e}}\psi$ and not $\psi$.

\section{Non-Abelian gauge theories}

Consider the pure Yang-Mills Lagrangian
\bea
{\cal {L}} &=& -{1\over 2} {\rm Tr}\left(F_{\mu\nu}F^{\mu\nu}\right)\,\,\,\,,
\nonumber \\
F_{\mu\nu} &=& \dd_\mu A_\nu -  \dd_\nu A_\mu -i g \left[A_\mu\,,\,A_\nu\right]\,\,\,\,.
\eea
Here $A_\mu=A_\mu^aT_a$ is a non-Abelian gauge field taking values, for example,  in the Lie algebra 
$SU(N)$ with commutation relations $\left[T_a\,,\,T_b\right]=
if_{ab}^cT_c$
and ${\rm Tr}(T_aT_b)={1\over 2}\delta_{ab}$.
This theory, when the gauge coupling $g$ is constant,  is invariant under
\be
A_\mu\longrightarrow hA_\mu h^\dagger - {i\over g } \dd_\mu hh^\dagger \,\,\,\,,
\ee
where the group element $h(x)$ is an arbitrary function in the Lie group
corresponding to the Lie algebra $SU(N)$.
\par
Let us now assume that the strength of the coupling between the gauge
fields is a space-time dependent quantity.  The Lagrangian we 
consider is given by 
\bea
{\cal {L}} &=& -{1\over 2} {\rm Tr}\left(F_{\mu\nu}F^{\mu\nu}\right)\,\,\,\,
\nonumber \\
F_{\mu\nu} &=& \dd_\mu A_\nu -  \dd_\nu A_\mu -i \widetilde{g} \left[A_\mu\,,\,A_\nu\right]\,\,\,\,,
\eea
where $\widetilde{g}=\widetilde{g}(x)$. We then demand that this Lagrangian is invariant under
the gauge transformation
\be
A_\mu\longrightarrow UA_\mu U^\dagger  - {i\over \widetilde{g} }  \dd_\mu U U^\dagger \,\,\,\,
\label{h-g}
\ee
for some Lie group element $U(x)$. 
Under this transformation, the field stength $F_{\mu\nu}$ transforms
as
\be
F_{\mu\nu}\longrightarrow U F_{\mu\nu}\,U^\dagger + {i\over \widetilde g^2}
\left(\dd_\mu\widetilde g\,\dd_\nu U U^\dagger -
\dd_\nu\widetilde g\, \dd_\mu UU^\dagger \right)\,\,\,\,.
\ee
If the field strength is to transforms as 
$F_{\mu\nu}\longrightarrow UF_{\mu\nu}\,U^\dagger $ (in order for  the
gauge kinetic term $-{1\over 2} {\rm Tr}\left(F_{\mu\nu}F^{\mu\nu}\right)$
to be invariant)  then  the condition
\be
\dd_\mu\widetilde g\, \dd_\nu U U^\dagger  -
\dd_\nu\widetilde g\,\dd_\mu U U^\dagger =0\,\,\,\,\,\,\,\,\,{\rm{or}}
\,\,\,\,\,\,\,\,\,d\widetilde g\wedge d UU^\dagger =0
\ee
must hold. This relation is satisfied 
provided that the Lie group element $U$ is an arbitrary function of $\widetilde{g}(x)$. That is, 
$U(x)=U(\widetilde{g}(x))$. Therefore, the non-Abelian gauge symmetry is not totally lost
if the gauge coupling $\widetilde{g}$ is not constant.

\par

The gauge coupling $\widetilde{g}$ characterises the 
interaction of the non-Abelian gauge fields between themselves and at the same time 
it describes the strength of the interaction of these gauge fields with any other fields. 
Let , for instance, 
$\Psi$ and $\bar\Psi$ 
be  a set  of fermions, carrying an index of the Lie Algebra
$SU(N)$, and 
transforming as
\bea
\Psi\longrightarrow U\psi\,\,\,\,\,,\,\,\,\,\,\,
\bar\Psi\longrightarrow \bar\Psi U^\dagger\,\,\,\,\,,
\eea
where we have suppressed the Lie algebra indices. 
The gauge covariant derivative
\be
{\cal D}_\mu\Psi =\left[\dd_\mu
-i \widetilde{g}(x) A_\mu \right]\Psi
 \,\,\,\,
\label{D}
\ee
transforms under (\ref{h-g}) as
\be
{\cal D}_\mu\Psi \longrightarrow U  \left({\cal D}_\mu \Psi\right)
\ee
and the Lagrangian
\bea
{\cal L} &=& -{1\over 2} {\rm Tr}\left(F_{\mu\nu}F^{\mu\nu}\right) 
+\bar \Psi\left[i\gamma^\mu{\cal D}_\mu - m\right]\Psi 
\nonumber \\
F_{\mu\nu} &=& \dd_\mu A_\nu -  \dd_\nu A_\mu -i\widetilde g(x) \left[A_\mu\,,\,A_\nu\right]
\,\,\,\,\,,
\eea
where the covariant derivative ${\cal D}_\mu$ is as defined in 
(\ref{D}), is gauge invariant.
This is the quantum chromodynamics Lagrangian with a space-time dependent
gauge coupling $\widetilde{g}$ (we have suppressed the Lie algebra indices 
in the second term).

\section{The Abelian Higgs model}

The Abelian Higgs model, 
with a space-time dependent gauge coupling, is described by the Lagrangian
\bea
{\cal L}_{U(1)} 
&=&
-\frac{1}{4}F_{\mu\nu}F^{\mu\nu}+
\left(\dd_\mu\phi^\star-i\widetilde eA_\mu\phi^\star\right)
\left(\dd^\mu\phi+i\widetilde eA^\mu\phi\right)
-m^2\phi^\star\phi
-\lambda\left(\phi^\star\phi\right)^2
\nonumber \\
F_{\mu\nu}&=&\dd_\mu A_\nu -\dd_\nu A_\mu\,\,\,\,\,\,.
\label{A-H-M}
\eea
The gauge coupling $\widetilde e$ is a space-time dependent function{\footnote{The parameters
$m^2$ and $\lambda$ could also be space-time dependent quantities. This does not affect
gauge symmetry.}}.
That is, $\widetilde e=\widetilde e(x)$. In this case, the Lagrangian
${\cal L}_{U(1)}$ is invariant under the gauge transformations
\be
A_\mu \longrightarrow A_\mu +\dd_\mu \alpha\,\,\,\,\,\,\,\,\,\,\,\,,\,\,\,\,\,\,\,\,\,
\phi \longrightarrow e^{-i\beta}\,\phi\,\,\,\,\,\,\,,
\ee
where $\alpha(x)$ is an arbitrary function of the gauge coupling $\widetilde e$.
That is $\alpha(x)=\alpha(\widetilde e(x))$ 
and $\beta$ depends on $\alpha$  through
\be
\beta({\widetilde{e}}(x))=\int \left({\widetilde{e}} {d\alpha\over d{\widetilde{e}}}\right)\, 
d{\widetilde{e}} 
={\widetilde{e}}\,\alpha\left({\widetilde{e}}\right)- \int \alpha\left({\widetilde{e}} \right)\, 
d\widetilde e 
\,\,\,\,\,\,\,\,.
\ee

\par
Let us now investigate the physical content of the above theory.  
We start by parametrising the complex scalar field $\phi$ as
\be
\phi=\rho\, e^{i\theta} \,\,\,\,.
\label{phi-para}
\ee
The gauge transformation $\phi \longrightarrow e^{-i\beta}\,\phi$ 
is now given by
\be
\theta \longrightarrow \theta -\beta
\ee
and $\rho(x)$ is unchanged as  $\rho^2= \phi^{*} \phi$ is gauge invariant.
With this parametrisation, the Lagrangian becomes
\bea
{\cal L}_{U(1)} &=&
-\frac{1}{4}F_{\mu\nu}F^{\mu\nu}
+
\widetilde e^2\rho^2\left(A_\mu +{1\over \widetilde e} \dd_\mu\theta 
+{i\over \widetilde e}\dd_\mu\ln\rho\right)
\left(A^\mu +{1\over \widetilde e} \dd^\mu\theta -{i\over\widetilde e}\dd^\mu\ln\rho\right)
\nonumber \\
&-& m^2\rho^2 - \lambda\rho^4
\,\,\,\,.
\label{rho-theta}
\eea

\par
In the Abelian Higgs model with constant gauge coupling,
the physical content is determined by choosing the unitary gauge
$\theta=0$. However, in the case when $\widetilde e$ is not constant,
the gauge $\theta=0$ cannot, in general, be reached. One atteins  the unitary gauge
by choosing the arbitrary function $\beta(\widetilde e(x))$ equal to $\theta(x)$ such that the transformed
field $\theta(x) -\beta(\widetilde e(x))$ vanishes. This is, in general, not possible as 
a function of four coordinates (that is $\theta(x)$) cannot be expressed in terms 
of a function of one variable only  (that is $\beta( \widetilde e(x))$ ). 

\par
If we insist on reproducing all the features of the Abelian Higgs 
model with constant gauge coupling, then we could demand that the 
field $\theta(x)$ is itself a function of $\widetilde e(x)$. That is,
\be
\theta(x)=\theta(\widetilde e(x))\,\,\,\,.
\ee
After all, $\theta$ is not a physical field (see below). In this case,
we could reach the unitary gauge $\theta(\widetilde e(x))=0$. 
We will, however, proceed in a way which is equivalent to choosing
the unitary gauge. It consists in working with gauge invariant variables.
Let us define the gauge invariant vector field
\be
V_\mu=A_\mu +{1\over \widetilde e} \dd_\mu\theta  \,\,\,\,.
\label{abelian-decomp}
\ee
Indeed, $V_\mu$ transforms as 
$V_\mu\longrightarrow V_\mu +\dd_\mu\alpha-{1\over \widetilde e}\,\dd_\mu\beta$
and $\dd_\mu\alpha-{1\over \widetilde e}\,\dd_\mu\beta=0$.
Notice also that the term ${1\over \widetilde e} \dd_\mu\theta$, 
if $\theta(x)=\theta(\widetilde e(x))$, can be written as $\dd_\mu \omega$,
where $\omega({\widetilde{e}}(x))=\int \left({1\over {\widetilde{e}}} 
{d\theta\over d{\widetilde{e}}}\right)d{\widetilde{e}} $.

\par

The Lagrangian of the Abelian Higgs model takes the form
\bea
{\cal L}_{U(1)}^{\rm{unitary}} &=&
-\frac{1}{4}V_{\mu\nu}V^{\mu\nu}+
\widetilde e^2\,\rho^2\, A_\mu A^\mu  +\dd_\mu\rho \dd^\mu\rho
-m^2\rho^2 - \lambda\rho^4
\nonumber \\
V_{\mu\nu} &=& \dd_\mu V_\nu - \dd_\nu V_\mu
\,\,\,\,\,.
\label{U-A-H-M}
\eea 
The field $\theta(\widetilde e(x))$ has disappeared and is, therefore, 
 not a true degree of freedom.
The mechanism of spontaneous symmetry breaking consists in expanding the
scalar field $\rho$ as 
\bea
\rho(x)=\rho_0+{\sigma(x)\over \sqrt{2}}) \,\,\,\,.
\eea
where 
the minimum
of the potential $V(\rho)=m^2\rho^2 + \lambda\rho^4$ is located at 
$\rho^2= \rho_0^2=-{m^2\over 2\lambda}$, with $m^2<0$ and $\lambda>0$.
The degrees of freedon
are therefore a massive vector field $V_\mu$ 
with a masse $M^2_V=2\rho_0^2 \widetilde e^2$
and a massive scalar field $\sigma$ (the Higgs field) with mass $M_\sigma^2=-m^2$.
These are precisely the properties of the Abelian Higgs model with constant gauge coupling.
However, the masse of the vector field $V_\mu$, 
when the gauge coupling is not constant, depends on space-time. 
We conclude that the mechanism of spontaneous symmetry breaking is, in this case, not
sufficient to guarantee a constant mass for the vector field $V_\mu$.

\section{The standard electro-weak theory}

The standard electro-weak theory, with non-constant gauge couplings,   
is described by the Lagrangian 
(see for instance \cite{ryder,quigg,zuber} for the case of constant gauge couplings)
\bea
{\cal L}_{SU(2)\times U(1)}={\cal L}_{{\rm{gauge}}} +  
{\cal L}_{{\rm{leptons}}} + {\cal L}_{{\rm{Higgs}}}
+ {\cal L}_{\rm{Yukawa}} \,\,\,\,.
\eea
The gauge part is 
\bea
{\cal L}_{{\rm{gauge}}} &=& -{1\over 2}{\rm Tr}\left(W_{\mu\nu} W^{\mu\nu}\right)
-{1\over 4}B_{\mu\nu} B^{\mu\nu}
\nonumber \\
W_{\mu\nu} &=& \dd_\mu {{W}}_\nu - \dd_\nu {{W}}_\mu
-i \widetilde{g}(x) \left[{{W}}_\mu \,,\,{{W}}_\nu \right]
\nonumber \\
B_{\mu\nu} &=& \dd_\mu B_\nu - \dd_\nu B_\mu
\,\,\,\,,
\eea
where the $SU(2)$ gauge field is ${{W}}_\mu=W^a T_a$ with 
the three matrices $T_a$ obeying 
the $SU(2)$ commutation relations $\left[T_a\,,\,T_b\right]=
i\epsilon_{abc}T_c$
and ${\rm Tr}(T_aT_b)={1\over 2}\delta_{ab}$. 
In the $2\times 2$ representation $T_a={1\over  2}\sigma_a$, where
$\sigma_a$ are the usual Pauli matrices.
The $U(1)$ gauge potential 
is denoted $B_\mu$.
\par
The $SU(2)$ gauge coupling $\widetilde{g}$ is taken to be a space-time 
dependent function, namely $\widetilde{g}=\widetilde{g}(x)$.
The $U(1)$ gauge coupling will be denoted $\widetilde{g}'=\widetilde{g}'(x)$ and is also
a space-time dependent quantity. 
\par
The Lagrangian ${\cal L}_{{\rm{gauge}}}$ is invariant under the gauge transformations 
\bea
W_\mu &\longrightarrow & UW_\mu U^\dagger -{i\over \widetilde{g}}\left(\dd_\mu U\right)U^\dagger
\nonumber \\
B_\mu  &\longrightarrow &  B_\mu + \dd_\mu \alpha
\eea
provided that the $SU(2)$ group element  $U$ (with $UU^\dagger =1$) is an arbitrary function
of $\widetilde{g}(x)$. That is,  $U= U(\widetilde{g}(x))$, as has been shown in section 5.

\par
The leptons (for simplicity, we include only the electron and its neutrino) enter through
\bea
{\cal L}_{{\rm{leptons}}}=i\bar R\gamma^\mu\left(\dd_\mu +i\widetilde{g}' B_\mu\right)R
+ i\bar L\gamma^\mu\left(\dd_\mu +{i\over 2} \widetilde{g}' B_\mu  
- i\widetilde{g} \, {{W}}_\mu \right)L \,\,\,\,
\eea
with 
\bea
L\equiv \left(\begin{array}{l} \nu_e \\ e_L\end{array}\right)
\,\,\,\,\,\,,\,\,\,\,\,\,
R\equiv e_R\,\,\,\,\,
\eea 
and for a fermion $\psi$ we have
$\psi_L={1\over 2}\left(1-\gamma_5\right)\psi$ and 
$\psi_R={1\over 2}\left(1+\gamma_5\right)\psi$. The left-handed neutrino 
is denoted $\nu_{\rm e}$ while ${\rm e}_L$ and ${\rm e}_R$
refer, respectively, to the left and the right chiralities
of the electron.
\par
When the $U(1)$ gauge coupling $\widetilde{g}'$ is a space-time dependent 
function, the fermions transform under the $U(1)$ gauge symmetry as
\bea
L\longrightarrow e^{-i\beta/2 }L \,\,\,\,\,\,\,,\,\,\,\,\,\,\,
R\longrightarrow e^{-i\beta }R 
\,\,\,\,,
\eea
where now the $U(1)$ gauge parameter $\alpha$ is an arbitrary function
of the $U(1)$ gauge coupling $\widetilde{g}'(x)$. 
That is, $\alpha=\alpha(\widetilde{g}'(x))$ and $\beta$ is given by    
\bea
\beta(\widetilde{g}')=\int \left(\widetilde{g}' {d\alpha\over d{\widetilde{g}}'}\right)\, 
d{\widetilde{g}}'\,\,\,\,\,\,
\eea
as has been established in section 4.

\par
On the other hand, under the $SU(2)$ gauge symmetry, 
the fermions transform as
\bea
L\longrightarrow U L \,\,\,\,\,\,\,,\,\,\,\,\,\,\,
R\longrightarrow  R 
\,\,\,\,.
\eea
Recall  that $U$ is a function of $\widetilde{g}(x)$.

\par
The complex scalar field 
\be
\Phi=\left(\begin{array}{c}
\phi_1\\ \phi_2\end{array}\right)
\ee
enters the electro-weak theory through the Lagrangian
\bea
{\cal L}_{{\rm{Higgs}}} &=& \left(D_\mu \Phi\right)^\dagger
\left(D^\mu \Phi\right)
-{m^2\over 2}\Phi^\dagger \Phi -{\lambda \over 4}\left(\Phi^\dagger \Phi\right)^2
\nonumber\\
D_\mu\Phi &=& \left(\dd_\mu - {i\over 2}\widetilde{g}'B_\mu - i \widetilde{g}W_\mu \right)\Phi
\,\,\,\,\,.
\eea
The complex scalar field has the $U(1)$ gauge symmetry
\be
\Phi \longrightarrow e^{i\beta/2} \Phi
\ee
and the $SU(2)$ gauge transformation
\be
\Phi\longrightarrow U\Phi\,\,\,\,\,.
\ee

Finally, the Yukawa part is given by
\bea
{\cal L}_{{\rm{Yukawa}}} =
-G_e\left(\bar L \Phi R +\bar R \Phi^\dagger L\right)\,\,\,\,,
\eea
where $G_e$ is the electron Yukawa coupling constant{\footnote{Gauge symmetry
does not prevent the parameters $m^2$, $\lambda$ and $G_e$ to be 
space-time dependent variables.}}.

\par
So far, we have shown that it is possible to render the gauge couplings 
of the standard electro-weak theory space-time dependent while  maintaining
some gauge symmetry. However, we still have to examine the spectrum of 
this theory. Let us recall that 
in the case of constant gauge couplings, the 
simplest way to get the spectrum is 
to choose for the scalar field $\Phi$ the unitary gauge
\be
\Phi=\left(\begin{array}{c}
0\\ \eta +{\sigma(x)\over \sqrt{2}}\end{array}\right)\,\,\,\,.
\label{higgs-gauge}
\ee
Here $\eta^2=-m^2/\lambda$, with $m^2<0$, is the 
ground state energy for the scalar field $\Phi$ (the minimum of the potential).

\par
On the other hand, for non-constant gauge couplings the above choice for the scalar
field is, in general, not atteinable. This is due to the fact that the $SU(2)$ gauge 
parameter is not an arbitrary function of space-time but an arbitrary
function of the gauge coupling $\widetilde{g}(x)$.  In other words, 
starting from the gauge choice (\ref{higgs-gauge}) one cannot reach
all the scalar field configurations by means of a gauge transformation
(it is, in general, not possible to ajust $\Phi(x)$ to a chosen gauge using 
a matrix $U(\widetilde g(x))$ that depends on $x^\mu$ only through $\widetilde g(x^\mu)$ ).
One way out of this is to assume that the the non-physical degrees of freedom
contained in the scalar field $\Phi$ are
a function of $\widetilde{g}(x)$.

\par
In order to see this, let us parametrise the scalar $\Phi$ as 
\be
\Phi=
\left(\begin{array}{c}
\phi_1\\ \phi_2\end{array}\right) 
=
\rho \left(\begin{array}{c}
\chi_1\\ \chi_2\end{array}\right)
\,\,\,\,\,,
\ee 
where $\rho(x)$ is defined as 
\bea 
\rho^2 = \Phi^\dagger \Phi=|\phi_1|^2+|\phi_2|^2\,\,\,\,\,.
\eea
The two complex fields $\chi_1$ and $\chi_2$ satisfy
\be
|\chi_1|^2+|\chi_2|^2=1\,\,\,\,\,
\ee
Since $\rho$ is a gauge invariant quantity, the $SU(2)$ gauge transformation
acts only on the fields $\chi_1$ and $\chi_2$. It is these fields
(that are not physical, as shown below) which we will
assume to depend on the the $SU(2)$ gauge coupling $\widetilde{g}(x)$.
Namely, 
\be
\chi_1(x)=
\chi_1(\widetilde{g}(x))\,\,\,\,\,\,\,\,\,\,,\,\,\,\,\,\,\,\,\,\,\,
\chi_2(x)=\chi_2(\widetilde{g}(x)) \,\,\,\,\,\,\,.
\ee 
In this way a matrix $U(\widetilde(g(x))$ can be found 
to reach  the unitary gauge (\ref{higgs-gauge}).  
We will, however,  choose to work with gauge invariant variables instead. 

\par

We start by noticing that given a vector
$\Phi=\rho\left(\begin{array}{c}
\chi_1\\ \chi_2\end{array}\right)$, such that  $|\chi_1|^2+|\chi_2|^2=1$,  one can 
always write it  as
\be
\Phi= \rho\,X^\dagger\,\left(\begin{array}{c}
0\\ 1\end{array}\right)\,\,\,\,\,,
\label{X-C}
\ee
where the matrix $X^\dagger$ 
belongs to the $SU(2)$ group ($XX^\dagger =1$ and $\det(X)=1$)
and is given by
\be
X^\dagger =\left(\begin{array}{cc}
\chi_2^{*} &\chi_1 \\
-\chi_1^{*} & \chi_2
\end{array}\right) 
\,\,\,\,\,.
\label{XX}
\ee 
The $SU(2)$ matrix $X^\dagger$ transforms as
\be
X^\dagger\longrightarrow UX^\dagger
\ee
in order for $\Phi$ to have the $SU(2)$ gauge transformation 
$\Phi\longrightarrow U\Phi$.

Our next step is to introduce the two variables 
\bea
W^X_\mu  &=&  X W_\mu X^\dagger - {i\over \widetilde g} \dd_\mu X X^\dagger
\nonumber \\
L^X &=&   XL
\label{change-var-unit}
\,\,\,\,\,.
\eea
The $U(1)$ gauge field  $B_\mu$ and fermionic singlet $R$ remain unchanged,
as they are not affected by the $SU(2)$ gauge transformation.
The $SU(2)$ vector field  $W^X_\mu$ and the fermionic doublet $L^X$ are 
gauge invariant under the $SU(2)$ gauge symmetry, as can be verified by using
the $SU(2)$ gauge transformations of $W_\mu$, $L$ and $X$.

\par
In order to find the expression of the electro-weak Lagrangian
in terms of the  new variables (this procedure is 
totally equivalent to choosing the unitary gauge
for which
$\Phi^X=\rho\left(
\begin{array}{c}
0 \\ 1
\end{array}
\right)=X\Phi\,\,$), 
we write the new gauge field $W_\mu^X$ as  
\be
W^X_\mu={1\over 2}\left({\cal W}_\mu^1\sigma_1+{\cal W}_\mu^2\sigma_2+{\cal W}_\mu^3\sigma_3\right)
={1\over 2}\left(\begin{array}{cc}
{\cal W}_\mu^3 & {\cal W}^-_\mu\\
{\cal W}^+_\mu  & -{\cal W}_\mu^3\end{array}
\right)
\,\,\,,
\label{w}
\ee
where we have defined the two vector fields 
\bea
{\cal W}_\mu^+ = {\cal W}_\mu^1 + i{\cal W}_\mu^2 
\,\,\,\,\,\,\,\,,\,\,\,\,\,\,\,\,
{\cal W}_\mu^- &=& {\cal W}_\mu^1 - i{\cal W}_\mu^2\,\,\,\,.
\eea

\par
The scalar sector in terms of the new variables is given by the Lagrangian
\bea
{\cal L}_{{\rm{Higgs}}}  
&=& \dd_\mu\rho\dd^\mu\rho 
+{1\over 4}\rho^2\left[\left(\widetilde g^2+\widetilde g'^2\right){\cal Z}_\mu {\cal Z}^\mu
+\widetilde g^2{\cal W}^+_\mu{\cal W}^-_\mu\right]
-{m^2\over 2}\rho^2 -{\lambda \over 4}\rho^4 
\,\,\,\,\,,
\eea
where we have also introduced the two gauge variables
\bea
{\cal A}_\mu &=& 
{\widetilde g' {\cal W}^3_\mu+{\widetilde g}B_\mu\over 
\left(\widetilde g^2+\widetilde g'^2\right)^{1/2}}
\nonumber \\
{\cal Z}_\mu &=& 
{\widetilde g {\cal W}^3_\mu-\widetilde g'B_\mu\over \left(g^2+\widetilde g'^2\right)^{1/2}}
\,\,\,\,
\eea
or,  equivalently
\bea
{\cal W}^3_\mu &=& 
{\widetilde g' {\cal A}_\mu+{\widetilde g}{\cal Z}_\mu\over 
\left(\widetilde g^2+\widetilde g'^2\right)^{1/2}}
\nonumber \\
B_\mu &=& 
{\widetilde g {\cal A}_\mu-\widetilde g'{\cal Z}_\mu\over \left(g^2+\widetilde g'^2\right)^{1/2}}
\,\,\,\,.
\label{Z-A}
\eea

\par

In order to find the expression of the fermionic Lagrangian
in terms of the new variables, we introduce the notation
\bea
L^X  = \left(\begin{array}{c}
{\cal N}_{\rm e} \\ {\cal E}_L
\end{array}\right)
\,\,\,\,\,,\,\,\,\,\,\,
R  = {\cal E}_R\,\,\,\,.
\eea
Explicitly, we have
\bea
{\cal L}_{{\rm{leptons}}}
&=& i\bar{\cal E}\gamma^\mu \left(\dd_\mu 
+i{\widetilde g\widetilde g'\over \left(\widetilde g^2
+\widetilde g'^2\right)^{1/2}} {\cal A}_\mu\right)
{\cal E} 
+i\bar{\cal N}_{\rm e}\gamma^\mu \dd_\mu {\cal N}_{\rm e}
\nonumber \\
&+& \left[{1\over 2}\left(\widetilde g^2+\widetilde g'^{2}\right)^{1/2}
\bar{\cal N}_{\rm e}\gamma^\mu {\cal N}_{\rm e}
+{\widetilde g'^{2}\over \left(\widetilde g^2+ \widetilde g'^{2}\right)^{1/2}}
\bar{\cal E}_R\gamma^\mu {\cal E}_R
- {\widetilde g^2- \widetilde g'^{2}\over 2
\left(\widetilde g^2+ \widetilde g'^{2}\right)^{1/2}}
\bar{\cal E}_L\gamma^\mu {\cal E}_L\right]{\cal Z}_\mu
\nonumber \\
&+&{\widetilde g\over 2}\left[\bar{\cal E}_L\gamma^\mu {\cal N}_{\rm e}{\cal W}_\mu^+
+ \bar{\cal N}_{\rm e}\gamma^\mu {\cal E}_L{\cal W}_\mu^-
\right]\,\,\,\,\,,
\eea 
where we have defined ${\cal E}={\cal E}_L+{\cal E}_R$.
\par
Similarly, the Yukawa part, in terms of the new variables,  yields
\bea
{\cal L}_{{\rm{Yukawa}}}^U  
&=& -G_{\rm e}\rho \left(\bar {\cal E}_L{\cal E}_R + \bar {\cal E}_R{\cal E}_L\right)
=-G_{\rm e}\,\rho\,\bar {\cal E} {\cal E}\,\,\,\,.
\eea

\par
The gauge part of the electro-weak theory is given by
\bea
{\cal L}_{{\rm{gauge}}} &=&
-{1\over 2}{\rm Tr}\left(W_{\mu\nu} W^{\,\mu\nu}\right)
-{1\over 4}B_{\mu\nu}B^{\mu\nu}
= -{1\over 2}{\rm Tr}\left(W^X_{\mu\nu} W^{X}_{\,\mu\nu}\right)
-{1\over 4}B_{\mu\nu}B^{\mu\nu}
\nonumber \\
W^X_{\mu\nu} &=& \dd_\mu {{W}}^X_\nu - \dd_\nu {{W}}^X_\mu
-i {\widetilde g}(x) \left[{{W}}^X_\mu \,,\,{{W}}^X_\nu \right]
\,\,\,\,\,.
\eea
The second equality holds because the matrix $X$ is, by assumption,  a function 
of $\widetilde g(x)$.
Using the expression of the matrix $W^X_\mu$ in (\ref{w}),
the field strength $W^X_{\mu\nu}$ takes the form
\bea
W^X_{\mu\nu}={1\over 2}\left(
\begin{array}{ccc}
{\cal W}_{\mu\nu}^3 -\widetilde gH_{\mu\nu} & 
\nabla_\mu {\cal W}^-_\nu -\nabla_\nu {\cal W}^-_\mu
\\
\nabla_\mu {\cal W}^+_\nu - \nabla_\nu {\cal W}^+_\mu &
-{\cal W}_{\mu\nu}^3 +\widetilde gH_{\mu\nu}
\end{array}\right)\,\,\,\,\,,
\eea
where 
\bea
{\cal W}^3_{\mu\nu} &=& 
\dd_\mu {\cal W}^3_\nu - \dd_\nu {\cal W}^3_\mu
\nonumber \\
H_{\mu\nu} &=& -{i\over 2}\left({\cal W}^+_\mu {\cal W}^-_\nu
-{\cal W}^+_\nu {\cal W}^-_\mu\right)
\nonumber \\
\nabla_\mu {\cal W}^+_\nu &=& \left(\dd_\mu 
+i\widetilde g{\cal W}^3_\mu\right){\cal W}^+_\nu
\nonumber \\
\nabla_\mu {\cal W}^-_\nu &=& \left(\dd_\mu 
-i\widetilde g{\cal W}^3_\mu\right){\cal W}^-_\nu
\,\,\,\,.
\eea

\par
Notice that one has terms involving $\dd_\mu W^3_\nu$ and $\dd_\mu B_\nu$
and upon replacing $W^3_\mu$ and $B_\mu$ by their expressions in  (\ref{Z-A}),
one generates quantities involving the derivatives of 
$\widetilde g(x)$ and  
$\widetilde g'(x)$.
Hence, if we want to have the same terms  as in the case 
of the standard electro-weak theory with constant gauge couplings
then we must demand that  
\be
{\widetilde{g} \over 
\left(\widetilde{g}^2 +\widetilde g'^{2}\right)^{1/2}}
=c\,\,\,\,\,\,\,\,\,\,\,,\,\,\,\,\,\,\,\,\,\,
{\widetilde g' \over 
\left(\widetilde{g}^2 +\widetilde g'^{2}\right)^{1/2}}
=c'
\,\,\,\,,
\label{assumption}
\ee
where $c$ and $c'$ are two constants. This means that
the two couplings $\widetilde g$ and $\widetilde g'$ are
related by $\widetilde g/\widetilde g'=c/c'$,
and one has only one space-time independent gauge coupling.

\par

By an explicit calculation, and using the assumption (\ref{assumption}),
we find that the gauge part is given by
\bea
{\cal L}_{{\rm{gauge}}} = 
&=& -{1\over 4}{\cal F}_{\mu\nu}{\cal F}^{\mu\nu}
-{1\over 4}{\cal Z}_{\mu\nu}{\cal Z}^{\mu\nu}
- {1\over 4}
\left(\nabla_\mu {\cal W}^+_\nu -\nabla_\nu {\cal W}^+_\mu\right)
\left(\nabla_\mu {\cal W}^-_\nu -\nabla_\nu {\cal W}^-_\mu\right)
\nonumber \\
&+& {1\over 2}{\widetilde g\over  \left(\widetilde g^2+ \widetilde g'^2\right)^{1/2}}
\left(\widetilde g{\cal Z}_{\mu\nu} +\widetilde g'{\cal F}_{\mu\nu}\right)H^{\mu\nu}
- {1\over 4}\widetilde g^2 H_{\mu\nu}H^{\mu\nu}
\,\,\,\,,
\label{gauge-explicit}
\eea
where
\bea
{\cal F}_{\mu\nu} &=& \dd_\mu {\cal A}_\nu - \dd_\nu {\cal A}_\mu
\nonumber \\
{\cal Z}_{\mu\nu} &=& \dd_\mu {\cal Z}_\nu - \dd_\nu {\cal Z}_\mu\,\,\,\,.
\,\,\,\,
\eea
are the field strenghts corresponding to the gauge fields ${\cal A}_\mu$
and ${\cal Z}_\mu$.

\par

To summarise, 
the full electro-weak
Lagrangian $
{\cal L}_{SU(2)\times U(1)}={\cal L}_{{\rm{gauge}}} 
+ {\cal L}_{{\rm{Higgs}}}+ {\cal L}_{{\rm{leptons}}}
+{\cal L}_{{\rm{Yukawa}}}$ with non-constant gauge couplings,
subject to the assumption (\ref{assumption}),   
contains the same terms 
as in the case of the electro-weak theory with constant gauge couplings. 
The fields $\chi_1$ and $\chi_2$
are unphysical as they have disappeared from the final theory 
(they have been absorbed by the non-Abelian gauge fields). 

\par
Furthermore, the Lagrangian ${\cal L}_{SU(2)\times U(1)}={\cal L}_{{\rm{gauge}}}
+ {\cal L}_{{\rm{Higgs}}}+ {\cal L}_{{\rm{leptons}}}
+{\cal L}_{{\rm{Yukawa}}}$ is still invariant under the $U(1)$ gauge symmetry
\bea
{\cal A}_\mu &\longrightarrow& {\cal A}_\mu +\dd_\mu\alpha
\nonumber \\
{\cal E} &\longrightarrow& e^{-i\beta}\,{\cal E}
\nonumber \\
{\cal W}^+_\mu &\longrightarrow& e^{-i\beta}\,{\cal W}^+_\mu 
\nonumber \\
{\cal W}^-_\mu &\longrightarrow& e^{i\beta}\,{\cal W}^-_\mu\,\,\,\,\,,
\eea
where $\alpha=\alpha(\widetilde e)$ is an arbitraty function of $\widetilde 
e={\widetilde g\widetilde g' \over 
\left(\widetilde{g}^2 +\widetilde g'^{2}\right)^{1/2}}$ and $\beta$ is given by
\be
\beta(\widetilde e)=\int\left(\widetilde e{d\alpha\over d\widetilde e}\right)
d\widetilde e\,\,\,\,.
\ee
The neutral fermion ${\cal N}_{\rm e}$ and the neutral vector field ${\cal Z}_\mu$
are not affected by this gauge symmetry. We should also mention that, according to
(\ref{Z-A}), ${\cal W}^3_\mu$ transforms as  ${\cal W}^3_\mu\longrightarrow{\cal W}^3_\mu 
+{\widetilde g' \over 
\left(\widetilde{g}^2 +\widetilde g'^{2}\right)^{1/2}}\dd_\mu\alpha$ leading to the
transformations $\nabla_\mu {\cal W}^+_\nu \longrightarrow e^{-i\beta}\,\nabla_\mu {\cal W}^+_\nu$
and $\nabla_\mu {\cal W}^-_\nu \longrightarrow e^{i\beta}\,\nabla_\mu {\cal W}^-_\nu$.
This shows that the Lagrangian ${\cal L}_{\rm{gauge}}$ in (\ref{gauge-explicit}) is 
explicitly gauge invariant.

The spectrum of the theory described by the full
Lagrangian 
$
{\cal L}_{SU(2)\times U(1)}={\cal L}_{{\rm{gauge}}}
+ {\cal L}_{{\rm{Higgs}}}+ {\cal L}_{{\rm{leptons}}}
+{\cal L}_{{\rm{Yukawa}}}$
is found by making the substitution
$\rho(x)=\eta +{\sigma(x)\over\sqrt{2}}$. 
It contains: $i)$ three massive vector fields
$\left({\cal Z}_\mu\,,\,{\cal W}_\mu^+\,,\,{\cal W}_\mu^-\right)$
and a massless gauge field ${\cal A}_\mu$.
$ii)$ A massive scalar field $\sigma$ (the Higgs field).
$iii)$ A massive fermion ${\cal E}$ (the electron) together with
a massless one ${\cal N}_{\rm e}$ (the neutrino). The latter does not couple 
to the massless vectror field  ${\cal A}_\mu$. The different masses 
are of course read from the quadratic parts of the Lagrangian.
However, the masses of the vector fields 
$\left({\cal Z}_\mu\,,\,{\cal W}_\mu^+\,,\,{\cal W}_\mu^-\right)$ are 
space-time dependent even after implementing the spontaneous symmetry
breaking mechanism.

\section{Conclusions}

It is commonly stated that non-constant gauge couplings are incompatible with gauge invariance.
We show in this paper that gauge invariance is not completely lost if the 
gauge couplings are not constant. This remark could be seen just as a mathematical curiosity
in its own right but it might also have some physical consequences especially in cosmology. 
\par
\par
It is certainly interesting to investigate the quantum properties of the various
gauge field theories presented in this paper. The simplest of these theories 
is obviously the one described by the Lagrangian 
\bea
{\cal L} = -{1\over 4}F_{\mu\nu}F^{\mu\nu} +\bar \psi\left[i\dirac
-{\widetilde{e}}(x)\gauge -m\right]\psi \,\,\,\,\,\,.
\eea
The first question to be asked is how to deal with the space-time dependent gauge coupling 
${\widetilde{e}}(x)$? We could regard ${\widetilde{e}}(x)$ as a dynamical field
having a Lagrangian of the form
\bea
{\cal L}_{{\widetilde{e}}} ={1\over 2}\dd_\mu{\widetilde{e}}\,\dd^\mu{\widetilde{e}}-
V\left(\widetilde{e}\right)\,\,\,\,\,,
\eea
where $V\left(\widetilde{e}\right)$ is some potential energy.
A dynamical field $\widetilde{e}$ might be  desirable from the point of view of cosmology and astrophysics
(if one includes gravity). However, it is problematic at the level of quantum field theory.
Indeed, the interaction term 
$\widetilde{e}\bar \psi \gauge\psi$ is , if $\widetilde{e}$ possesses a kinetic term, 
a dimension five operator and leads to a non-renormalisable theory.
\par
On the other hand, if the Lagrangian  ${\cal L}_{{\widetilde{e}}}$ is not included
then one could view the non-constant gauge coupling ${\widetilde{e}}(x)$
as a non-propagating  background. The Feynman rules and the Feynman graphs are then exactly those of quantum electrodynamics 
with constant gauge coupling. In this case, we expect the theory to be renormalisable.

\end{document}